 \documentclass[12pt,preprint]{aastex}    
 \usepackage{emulateapj5}

\newcommand{\gsim}{\;\lower4pt\hbox{${\buildrel\displaystyle >\over\sim}$}\;}
\newcommand{\lsim}{\;\lower4pt\hbox{${\buildrel\displaystyle <\over\sim}$}\;}

\font\eightrm=cmr8 scaled 1000

\slugcomment{To appear in the Astrophysical Journal Letters }
\shorttitle{MHD tidal waves on spinning compact objects}
\shortauthors{Lou}

\begin{document}

\title{MHD tidal waves on a spinning magnetic compact star}


\author{Yu-Qing Lou \altaffilmark{1,\hbox{ }2} }

\vskip -0.2cm
\affil{\vskip 0.4cm $^1$National Astronomical Observatories,
Chinese Academy of Sciences, A20, Datun Road, 
%Chaoyang, 
Beijing, 100012 }
%
%\and
%
\affil{$^2$Department of Astronomy and Astrophysics, 
    The Univ. of Chicago,
    5640 S. Ellis Ave, Chicago, IL 60637 }
\email{lou@oddjob.uchicago.edu}
%
%referee report on August 16, 2001 MS#15530
%September 1, 2001 Beijing
%November 1, 2001 the second referee report in apjl07.2rpt


\begin{abstract}
In an X-ray binary system, the companion star feeds the compact neutron 
star with plasma materials via accretions. The spinning neutron star is 
likely covered with a thin ``magnetized ocean" and may support 
{\it magnetohydrodynamic (MHD) tidal waves}. While modulating the 
thermal properties of the ocean, MHD tidal waves periodically shake 
the base of the stellar magnetosphere that traps energetic particles, 
including radiating relativistic electrons. For a radio pulsar, MHD 
tidal waves in the stellar surface layer may modulate radio emission 
processes and leave indelible signatures on timescales different 
from the spin period. 
%For a neutron star that spins sufficiently fast, gravitational 
%radiations may induce instabilities for such MHD tidal waves. 
Accretion activities are capable of exciting these waves but may also 
obstruct or obscure their detections meanwhile. Under fortuitous 
conditions, MHD tidal waves might be detectable and offer valuable 
means to probe properties of the underlying neutron star. Similar 
situations may also occur for a cataclysmic variable -- an accretion 
binary system that contains a rotating magnetic white dwarf. This 
Letter presents the theory for MHD tidal waves in the magnetized 
ocean of a rotating degenerate star and emphasizes their potential 
diagnostics in X-ray and radio emissions. 
%As a possible case, we suspect that small-amplitude pulsations in 
%x-ray flux from the LMXB system XTEJ1808-369 might be MHD tidal 
%wave signals.
\end{abstract}



%% chosen the first three characters of the first author's name 
%% plus the last two numeral of the year of publication as our 
%% KEY for each reference.

\section{Introduction}

For a magnetized spinning degenerate compact star covered 
with a thin plasma layer referred to as an {\it ocean}, various 
magnetohydrodynamic (MHD) tidal waves, excited by relevant 
processes, may exist. This scenario is plausible for an 
accretion binary system in which the Roche-overflow from a 
companion star accretes onto a spinning magnetized compact 
star. This Letter describes the theory of MHD tidal waves that 
propagate in a thin magnetized ocean of a spinning compact star. 
%For observational prospects, we note the following. 1. 
When combined with pertinent dynamic processes and emission 
mechanisms, such MHD tidal waves may leave indelible signatures 
of electromagnetic emissions in a wide spectrum ranging from
the radio to GeV gamma rays
%(e.g., radio, optical, and x-ray etc.) 
and could provide key diagnostics that might offer 
clues and hints of otherwise complex phenomena.
%2. For wave excitation, there are instabilities induced by 
%gravitational radiations (Andersson et al. 1999) given a 
%specific situation. Or, 
In general, MHD tidal waves may be excited gravitationally,
thermally, electromagnetically, or by impacts of accreting 
materials. While accretion activities may stir up and sustain 
these waves, wave signals might be undesirably entangled with or 
overwhelmed by accretion noises. Once observationally established, 
these waves may be utilized to probe the structures in the 
{\it surface layers} of neutron stars and white dwarfs by 
extending or adapting the helioseismological techniques 
(Papaloizou \& Pringle 1978; van Horn 1980; Saio 1982; Lou 
2000, 2001b). Furthermore, damping properties of these waves 
might provide information of the stellar constituent materials 
and diffusion processes (Lou 1995).

\section{Theoretical Model Analysis and Results}

MHD tidal waves may exist on spinning magnetic degenerate stars 
in either isolation or binaries (Chanmugam 1992; Phinney \& 
Kulkarni 1994; Bildsten \& Strohmayer 1999), although they might
have a better chance to be detected in high-mass and low-mass 
x-ray binaries (HMXBs and LMXBs) or cataclysmic variables. By 
strong gravity, the magnetized plasma ocean 
%(Brown \& Bildsten 1998) 
on a degenerate star of radius $R$ is thin. As the ratio 
$eR^3\Omega B/(GMm_pc)$ of electric force to surface gravity can 
be large for either neutron stars or white dwarfs in rotation 
(Goldreich \& Julian 1969), where $e$ is the electron charge, 
%$R$ is the stellar radius, 
$\Omega$ is the stellar angular spin rate, $B$ is the poloidal 
magnetic field, $c$ is the speed of light, $G$ is the gravitational 
constant, $M$ is the stellar mass and $m_p$ is the proton mass, 
excess surface charges will be pulled upwards against gravity 
to form stellar magnetospheres. It is presumed that the ocean and 
an extended magnetosphere coexist on a compact star, that the 
magnetosphere maintains a pressure at the interface between the 
ocean and the magnetosphere much as the Earth's atmosphere does 
on the ocean (Gill 1982), and that magnetic fields (idealized
as a split monople) are in and out radially across the thin ocean 
of thickness $h$. 
%and the stellar surface is covered with a few 
%adjacent patches of opposite magnetic polarities. 
For a zonal magnetic field, the Alfv\'en-Rossby waves have 
been studied earlier in the solar/stellar context (Gilman 
1969; Lou 1987, 2000).

For $h/L\ll 1$ and a small Rossby number 
${\cal R}\equiv U/(2\Omega L)$ where $U$ and $L$ are 
typical horizontal velocity and spatial scales,
%($h$ is the ocean depth), 
the shallow-magnetofluid approach is valid. 
With a quasi-static vertical balance, 
the MHD tidal equations in spherical 
polar coordinates $(r,\theta,\phi)$ are:
$$
{\partial v_{\theta}\over\partial t}-f v_{\phi}
=-{1\over R}{\partial\Theta\over\partial\theta}
%\bigg(g\eta+{Bb_r\over 4\pi\rho}\bigg)
-{C_A^2\over c^2}{\partial v_{\theta}\over\partial t}\ ,\eqno(1)
$$
$$
{\partial v_{\phi}\over\partial t}+f v_{\theta}
=-{1\over R\sin\theta}{\partial\Theta\over\partial\phi}
%\bigg(g\eta+{Bb_r\over 4\pi\rho}\bigg)
-{C_A^2\over c^2}{\partial v_{\phi}\over\partial t}\ ,\eqno(2)
$$
$$
{\partial\eta\over\partial t}+{h\over R\sin\theta }
\bigg[{\partial\over\partial\theta }(\sin\theta v_{\theta})+
{\partial v_{\phi}\over\partial\phi}\bigg]=0\ ,\eqno(3)
$$
$$
{\partial b_r\over\partial t}+{B\over R\sin\theta}
\bigg[{\partial\over\partial\theta }(\sin\theta v_{\theta})+
{\partial v_{\phi}\over\partial\phi}\bigg]=0\ ,\eqno(4)
$$
where variables $v_{\theta}$, $v_{\phi}$, $\eta$ and $b_r$ are the 
perturbations of $\hat\theta-$component velocity, $\hat\phi-$component 
velocity, ocean surface elevation, and $\hat r-$component of magnetic 
field, respectively, 
$f\equiv 2\Omega\cos\theta$, $g\equiv GM/R^2$, $B$ and $\rho$ are 
the Coriolis parameter at colatitude $\theta$, the surface gravity, 
the radial magnetic field, and the mean mass 
density, respectively, $C_A\equiv B/(4\pi\rho)^{1/2}$ is the Alfv\'en 
speed, $\Theta\equiv g\eta +Bb_r/(4\pi\rho)$ is
proportional to the total pressure perturbation,
and the {\it displacement current} effect is retained in 
case of $C_A\gsim c$ (Lou 1995). The energy conservation 
follows from eqns. (1)-(4) with the energy flux density
$\vec v_{\perp}[\rho g\eta+Bb_r/(4\pi)]$
%\begin{eqnarray}
%& & {\partial\over\partial t}\bigg[\bigg(1+{C_A^2\over c^2}\bigg)
%{\rho h(v_{\theta}^2+v_{\phi}^2)\over 2}+{\rho g\eta^2\over 2}
%+{hb_r^2\over 8\pi}\bigg] \nonumber \\
%& &\qquad\qquad\qquad\quad
%+\nabla_{\perp}\cdot\bigg[\rho h\vec v_{\perp}
%\bigg(g\eta+{Bb_r\over 4\pi\rho}\bigg)\bigg]=0\ ,\quad\ \  \nonumber (5)
%\end{eqnarray}
where $\vec v_{\perp}\equiv (v_{\theta}, v_{\phi})$. By the 
$\exp[i(\omega t+m\phi)]$ dependence with angular frequency 
$\omega$ and integer $m$, the resulting equation of 
$\Theta(\mu)$ becomes 
\begin{eqnarray}
& & \!\!\!\!
{d\over d\mu}\bigg({1-\mu^2\over s^2-\mu^2}
{d\Theta\over d\mu}\bigg)-{1\over s^2-\mu^2}
\bigg({m\over s}{s^2+\mu^2\over s^2-\mu^2}
+{m^2\over 1-\mu^2}\bigg)\Theta \nonumber \\
& &\quad\qquad\qquad\qquad\quad
+{4R^2\Omega^2\Theta\over (gh+C_A^2)
(1+C_A^2/c^2)}=0 ,\qquad\ \nonumber (5)
\end{eqnarray}
where $\mu\equiv\cos\theta$ and
$s\equiv\omega (1+C_A^2/c^2)/(2\Omega)$.
%$\Omega$ is the stellar angular spin rate. $R$ the stelar radius, 
%$g$ the surface gravity, and $h$ is the thin ocean layer thickness.
%(see, e.g., equation (51) of Chapman \& Lindzen 1970 on page 113).
Variables $v_{\theta}$, $v_{\phi}$, $\eta$ and $b_r$ 
may be explicitly expressed in terms of $\Theta(\mu)$.
%$$
%v_{\theta}={is\exp[i(\omega t+m\phi)]
%\over 2\Omega R(s^2-\cos^2\theta )}
%\bigg({d\over d\theta}+{m\cot\theta\over s}\bigg)\Theta\ ,\eqno(7)
%$$
%$$
%v_{\phi}={-s\exp[i(\omega t+m\phi)]
%\over 2\Omega R(s^2-\cos^2\theta )}
%\bigg({\cos\theta\over s}{d\over d\theta}+
%{m\over\sin\theta}\bigg)\Theta\ ,\eqno(8)
%$$
%$$
%\eta = {h\Theta\exp[i(\omega t+m\phi)]\over gh+B^2/(4\pi\rho)}\ ,\eqno(9)
%$$
%$$
%b_r={B\Theta\exp[i(\omega t+m\phi)]\over gh+B^2/(4\pi\rho)}\ .\eqno(10)
%$$
The $\Theta (\mu)$ solutions, regular at the two poles, involve the 
{\it Hough functions} (Hough 1897, 1898; Flattery 1967; Longuet-Higgins 
1968; Chapman \& Lindzen 1970). Given relevant physical parameters, all 
properties of MHD tidal waves can be derived computationally.

Without magnetic field, eqn (5) is 
equivalent to eqns (7) and (8) of Bildsten et al.
%, Ushomirsky, \& Cutler 
(1996) with their two parameters $q^2\equiv s^{-2}$ and 
$\lambda\equiv R^2\omega^2/(gh)$. In their ``traditional 
approximation", the thin compressible layer is disturbed 
adiabatically, the radial component of the Coriolis force 
is ignored, the radial displacement is much less than the 
horizontal displacement, and $\omega$ is much lower than 
the Brunt-V\"ais\"al\"a frequency. 
%Their $\lambda$ (BUC1996) was chosen to be $l(l+1)$ for 
%integers $l$ so that $q$ becomes an eigenvalue parameter. 
They focus on internal gravity modes of a rotating ocean. While 
internal magneto-gravity modes (Lou 1996) are excluded by invoking 
the quasi-static vertical balance, our incompressible formulation 
contains various MHD tidal waves in different regimes associated 
with the distinguished names of Kelvin (1879), Alfv\'en (1942), 
Rossby (1939), and Poincar\'e (1910), leading to the acronym of 
KARP waves.

For analytical simplicity, we hereafter invoke the $\beta-$plane 
approximation (Rossby et al. 1939) that retains the essential 
physics in the spherical geometry. In the local $\beta-$plane 
approximation of eqns $(1)-(4)$ with 
$\beta\equiv 2\Omega\sin\theta/R$, a Cartesian coordinate system 
$(x,y,z)$ is set up so that $\hat x$, $\hat y$ and $\hat z$ point 
eastward (the sense of spin), northward and radially outward, 
respectively. One derives
$$
\bigg[
{\partial\over\partial t}\bigg({1\over c^{*2}}
{\partial^2\over\partial t^2}+{f^{*2}\over c^{*2}}
-{\partial^2\over\partial x^2}
-{\partial^2\over\partial y^2}\bigg)
-\beta^{*}{\partial\over\partial x}\bigg]
v_{\theta}=0\ ,\eqno(6)
$$
where $f^{*}\equiv f/(1+C_A^2/c^2)$ is the {\it modified 
$f-$parameter}, $\beta^{*}\equiv\beta/(1+C_A^2/c^2)$ is 
the {\it modified $\beta-$parameter}, and $c^{*2}\equiv 
(gh+C_A^2)/(1+C_A^2/c^2)$ defines a characteristic speed 
$c^{*}$. For free MHD tidal waves of {\it small} 
$y-$variation scales and with 
$v_{\theta}\propto\exp(i\omega t-ik_xx-ik_yy)$ where $k_x$ 
and $k_y$ are $x-$ and $y-$component wavenumbers, the cubic 
dispersion relation is 
$
\omega^3-[f^{*2}+c^{*2}(k_x^2+k_y^2)]
\omega-\beta^{*} c^{*2}k_x=0.
$                                                                          
Global versions of KARP waves are contained in eqn (5). 
%------------------------------------------------------------
%$$
%\omega^3-\bigg[{f^2\over (1+C_A^2/c^2)^2}
%+c_R^2(k_x^2+k_y^2)\bigg]\omega-{\beta c_R^2k_x
%\over (1+C_A^2/c^2)}=0\ \eqno(5)
%$$
%-------------------------------------------------------------

\section{Equatorially Trapped MHD Tidal Waves}

By a latitudinal variation of $f=\beta y$, the 
{\it equatorial $\beta-$plane} becomes an effective 
{\it waveguide} for trapping, within the equatorial zone, 
MHD tidal waves that propagate in the $\hat\phi-$direction 
and this equatorial trapping effect becomes stronger in 
the regime of larger $4R^2\Omega^2/[(gh+C_A^2)(1+C_A^2/c^2)]$ 
in eqn (5) (Longuet-Higgins 1968). 
%These MHD tidal waves are latitudinally confined to 
%the equatorial zone and propagate in the azimuthal 
%$\hat\phi-$direction (e.g., Gill 1982; Pedlosky 1987). 
We examine main properties of such {\it equatorially 
trapped MHD tidal waves} below.

For {\it equatorially trapped Alfv\'en-Kelvin waves} 
(Thompson 1879; Alfv\'en 1942) around $\theta=\pi/2$ 
that propagate only {\it eastward} at speed $c^{*}$, 
one must directly solve eqns $(1)-(4)$ with 
$v_{\theta}=0$ to obtain
$$
\!
\{v_{\phi},\eta, b_r\}=\{c^{*}, h, B\}
\exp[-\beta^{*}y^2/(2c^{*})]
{\cal G}(x-c^{*}t) ,\eqno(7)
$$
where ${\cal G}(\cdots)$ is a {\it small arbitrary} 
dimensionless waveform and $\beta^{*}$ appears in 
the Gaussian $y-$profile that decays at large $|y|$ 
(e.g., Gill 1982).

With $f=\beta y$ and a {\it nonzero}
$v_{\theta}\propto\exp(ik_xx-i\omega t)$ in the equatorial 
zone around $\theta=\pi/2$, eqn (6) becomes the Schr\"odinger 
equation of a harmonic oscillator. The normalized real 
solution takes the form of 
$$
v_{\theta}={\cal N}
\exp\bigg[-{\beta^{*}y^2\over 2c^{*}}\bigg]
H_n[(\beta^{*}/c^{*})^{1/2}y]
\cos(k_xx-\omega t)\ \eqno(8)
$$
with $H_n(\xi)$ being a Hermite polynomial 
of order $n$ and the normalization constant
${\cal N}\equiv 2^{-n/2}(n!)^{-1/2}
[\beta^{*}/(\pi c^{*})]^{1/4}$. Variables
$v_{\phi}$, $\eta$, and $b_r$ can be derived 
accordingly. The resulting {\it dimensionless} 
cubic dispersion relation for {\it equatorially 
trapped KARP waves} becomes 
$$
\tilde\omega^3-(\tilde k_x^2+n+1/2)\tilde\omega
-\tilde k_x/2=0\ , \eqno(9)
$$
%$$
%\omega^2/c^{*2}-k_x^2-\beta^{*}k_x/\omega=
%(2n+1)\beta^{*}/c^{*}\ \eqno(7)
%$$
%$$
%{\omega^2\over c_R^2}-k_x^2-{\beta k_x\over\omega
%(1+C_A^2/c^2)}={(2n+1)\beta\over c_R(1+C_A^2/c^2)}\ \eqno(7)
%$$
where $\tilde\omega\equiv\omega/(2\beta^{*}c^{*})^{1/2}$,
$\tilde k_x\equiv k_x/(2\beta^{*}/c^{*})^{1/2}$, and integer 
$n\geq 0$ gives the number of nodes along $y-$direction. Results of 
eqn (9) are displayed in Figure 1 with the straight dashed line 
for the Alfv\'en-Kelvin mode (7) {\it added}. For $n\geq 1$, 
there are two classes of MHD tidal waves. The high-frequency 
class (upper solid lines) corresponds to {\it equatorially 
trapped Alfv\'en-Poincar\'e waves} (Poincar\'e 1910; Alfv\'en 
1942) that propagate either eastward or westward but with 
{\it different} phase speeds $\omega/k_x$ for a given $|k_x|$. 
The low-frequency class (lower solid lines) corresponds to 
{\it equatorially trapped Alfv\'en-Rossby waves} (Rossby et al. 
1939; Alfv\'en 1942) with westward phase speeds; the group 
velocity $d\omega/dk_x$ is eastward and westward for short- 
and long-waves, respectively. The $n=0$ case (dash-dot line) 
of eqn (9) corresponds to
$
\tilde\omega=\tilde k_x/2\pm 
(\tilde k_x^2/4+1/2)^{1/2}
$
and is referred to as {\it equatorially trapped mixed 
Alfv\'en-Rossby-Poincar\'e waves} (Poincar\'e 1910; Rossby et al. 
1939; Alfv\'en 1942) whose phase velocity can be to the east 
(plus sign) or west (minus sign) yet with a group velocity 
always to the east; for large $k_x>0$ it behaves more like 
Alfv\'en-Poincar\'e waves (plus sign), while for large $|k_x|$ 
with $k_x<0$ it behaves more like Alfv\'en-Rossby waves (minus 
sign). There are no standing modes (i.e., $\tilde k_x=0$)
for equatorially trapped Alfv\'en-Kelvin and Alfv\'en-Rossby 
waves (Fig. 1). In contrast, for equatorially trapped 
Alfv\'en-Poincar\'e waves and mixed Alfv\'en-Rossby-Poincar\'e 
waves, {\it axisymmetric standing modes} exist with 
$\tilde k_x=0$ in eq. (9) and thus $\tilde\omega^2=n+1/2$ where 
$n=0,1,2,\cdots$.

%\vskip 7.0cm
%\figcaption[]{Dispersion relations of KARP or MHD tidal waves.}

\section{Estimates and Diagnostics}

The preceding analysis is succinct with potential applications 
to diverse stellar settings. The estimates here focus on typical 
neutron stars. One can extend the analysis to white dwarfs with 
different parameters. With $M\sim 1.4M_{\odot}$ and $R\sim 10^6$cm, 
the surface gravity of a neutron star is 
$g\sim 10^{14}\hbox{cm s}^{-2}$. The ocean surface gravity wave 
speed is $(gh)^{1/2}\sim 10^9h_4^{1/2}\hbox{cm s}^{-1}$ and the 
Alfv\'en speed is $C_A\sim 10^9B_{12}\rho_5^{-1/2}\hbox{cm s}^{-1}$ 
for HMXBs or $C_A\sim 10^6B_{9}\rho_5^{-1/2}\hbox{cm s}^{-1}$ for 
LMXBs, where $h_{\delta}$, $\rho_5$, and $B_{\alpha}$ stand for $h$, 
$\rho$, and $B$ in units of $10^{\delta}$cm, $10^5$g, and 
$10^{\alpha}$G, respectively. For an extreme magnetar of 
$C_A\sim 10^{12}B_{15}\rho_5^{-1/2}\hbox{cm s}^{-1}$, relativistic 
and quantum effects are involved. For spin periods ranging from 
$\sim 16$ms to $\sim 8.5$s, the rotation speed $\Omega R$ ranges 
from $\sim 10$ to $\sim 10^4\hbox{ km s}^{-1}$. 
It is clear that MHD tidal waves are important in 
the ocean dynamics of a spinning neutron star.

For a magnetar or neutron star in a HMXB, the situation of 
$C_A\gsim c$, $\Omega R$ and $(gh)^{1/2}$ may happen with a 
stronger $B$ and a less dense ocean. The inclusion of the 
displacement current term in eqns (1), (2), (5) and (6) is 
necessary (see the definitions of $f^*$, $\beta^*$, and 
$c^*$ following eq [6]) and the magnetic field is dynamically 
significant in MHD tidal waves besides potential diagnostic 
roles.

For a neutron star in a LMXB,
%of XTE J1808-369 (Wijnands \& van der Klis 1998; Chakrabarty 
%\& Morgan 1998; Psaltis \& Chakrabarty 1999), one would have 
$C_A\ll c$, $C_A<(gh)^{1/2}$, and $C_A\cong\Omega R$ unless 
$B_{9}\gg 1$, $\rho_{5}\ll 1$ and $h_2\ll 1$. Thus, $B$ 
affects only the rotational contribution to the speeds and 
frequencies of tidal waves significantly. Flow velocities of 
KARP waves are powerful to push magnetic fields around as 
$v_{\theta}^2+v_{\phi}^2>b_r^2/(4\pi\rho)$. Meanwhile, a $B$ 
of $\sim 10^{8}-10^9$ G would be strong enough to partially 
guide materials from an accretion disk onto the magnetic poles 
(White et al. 1983), leading to intense X-ray emissions as 
well as sporadic bursts (Bildsten \& Strohmayer 1999). 

KARP waves might be driven to excitation by radiative thermal or 
nuclear processes, by polar accretion impacts, by magnetic thrust 
and dragging, or by tidal interactions of the accretion disk,
%\footnote{The tidal elevation $|\eta|$
%due to a companion star is 
%$|\eta|\sim (m_2/M)(R/\Lambda)^3R$ approximately, where $m_2$ 
%is the companion mass and $\Lambda$ is the distance between 
%the two stars. For the inferred parameters of XTEJ1808-369 as 
%an example, $m_2/M\sim 0.1$ and $R/\Lambda\sim 3\times 10^{-5}$ 
%(Wijnands \& van der Klis 1998; Chakrabarty \& Morgan 1998). 
%For the gravitational tidal effect of $m_2$ {\it alone}, 
%$|\eta|\sim 10^{-9}$ cm is negligible.} 
and in turn, are capable 
of periodically modulating thermal properties of the ocean and 
buffeting magnetic fields in polar regions to affect the height 
and cross section of polar accretions. 
%Through magnetospheric 
%wave transmissions, KARP waves and polar accretion activities 
%might also influence the inner region of a circumstellar
%accretion disk.

As an example of illustration, we now discuss 
possible diagnostics for equatorial MHD tidal waves of a neutron 
star in a LMXB. In spherical geometry, $k_x=m/R$ for an integer
%\footnote{The stellar circumference is $2\pi R=m\Sigma$ with $\Sigma$ 
%being the azimuthal wavelength. The local-to-global wavenumber extension 
%from $k_x$ to $m/R$ remains {\it rigorously valid} for the azimuthal 
%dimension in the spherical geometry (Longuet-Higgins 1968).} 
$m$ and $\tilde k_x\sim 0.18m(M_{33}h_4)^{1/4}$ $(R_6\nu_{3})^{-1/2}$ 
for a {\it small} $C_A$, where $\nu_3$, $M_{33}$, and $R_6$ stand for
$\nu\equiv\Omega/(2\pi)$, $M$, and $R$ in units of $10^3$Hz, $10^{33}$g, 
and $10^6$cm, respectively. 
%$h_4$ is the plasma ocean depth $h$ in units 
%of $10^4$ cm, (Brown \& Bildsten 1998) 
The Rossby radius of deformation 
%(Gill 1982)
is $D_{rossby}\equiv ({2\beta^{*}/c^{*}})^{-1/2}$
$\sim 1.8\times 10^5(M_{33}h_{4})^{1/4}\nu_{3}^{-1/2}\hbox{ cm}$.
For $R_6\sim 1$, $M_{33}\sim 2.8$ 
%($1.4M_{\odot}$) 
and $\nu_3\sim 0.5$, equatorially trapped KARP waves influence 
a latitude range $\gsim\pm 20^{\circ}$ across the equator. 
%That is, the KARP wave amplitude drops to $\sim 0.78$ of its equatorial 
%value at $D_{rossby}$ north or south of the equator, or to $\sim 0.37$ 
%at 2$D_{rossby}$ (a latitude span of $\sim\pm 40^{\circ}$ in the latter 
%case). The scenario for XTE J1808-369 might be as follows. (Wijnands 
%\& van der Klis 1998; Chakrabarty \& Morgan 1998)

Take a plausible geometry favorable for a detection. The spin and 
magnetic axes are misaligned with the two magnetic poles 
at latitudes outside $\pm 45^{\circ}$.
The spin axis, likely aligned with the binary orbital axis, is
%(an orbital period of two hours)
taken to be perpendicular to the line of sight.
%(Chakrabarty \& Morgan 1998).
%(cf. ref. {\it (2)} for considerations of possible binary inclinations).
During the phase of moderate and steady polar accretions from the
surrounding disk, the stellar surface gives off a largely 
isotropic X-ray flux sufficiently away from the two magnetic 
polar hot spots. In this scenario,
%(Chakrabarty \& Morgan 1998), 
one may not directly detect either X-ray hot spots. {\it Instead, 
accretion-powered KARP waves trapped in the equatorial belt may 
give rise to relatively smooth large-scale ($L\gsim R$) periodical 
modulations of X-ray brightness}. In principle, all these waves may 
coexist. Estimates are given below for $h_4\sim 1$, $R_6\sim 1$, 
and $M_{33}\sim 2.8$.
%as an exercise.

For $m=0$ or $\tilde k_x=0$, dispersion relation (9) 
gives {\it standing} MHD tidal wave frequencies 
$f_{\hbox{\eightrm w}}$ as 
\begin{eqnarray}
& &\!\!\!
 f_{\hbox{\eightrm w}}
=\pm{(2n+1)^{1/2}\over 2\pi}
\bigg[{2\Omega\over R}{(GMh/R^2+C_A^2)^{1/2}\over
(1+C_A^2/c^2)^{3/2}}\bigg]^{1/2}\nonumber \\
& &\ \ \cong\pm 5.1\times 10^2(2n+1)^{1/2}\, \hbox{ }
\nu_3^{1/2}(M_{33}h_4)^{1/4}R_6^{-1}\,\,\hbox{Hz}\ \ \nonumber\ (10)
\end{eqnarray}
with $n=0,1,2,\cdots$. Given a detected $f_{\hbox{\eightrm w}}$, 
the stellar spin rate of $\nu$ 
%($\Omega\equiv 2\pi\nu$) 
can be inferred for $n=0$ -- the special case of mixed 
Alfv\'en-Rossby-Poincar\'e standing waves (the dash-dot line 
in Fig. 1 with $k_x=0$). Or, for standing Alfv\'en-Poincar\'e 
waves of larger $n\geq 1$, $\nu$ would be smaller. 

For nonaxisymmetric waves, the frequencies of 
equatorially trapped Alfv\'en-Kelvin waves (7) (the 
dashed straight line of Fig. 1) seen in an inertial 
reference frame are
$$
f_{\hbox{\eightrm w}}=10^3m\nu_3
+1.3\times 10^2m(M_{33}h_4)^{1/2}R_6^{-2}\quad\hbox{Hz}\ \eqno(11)
$$
for a {\it small} $C_A$, where the corotation effect is included. With 
$m=1$ and a detected $f_{\hbox{\eightrm w}}$, one can infer $\nu$. 
For $m=2$ and 3, $f_{\hbox{\eightrm w}}$ would be {\it precisely} 
2 and 3 times the fundamental frequency.
%\footnote{By equations (16) and (22), a weak $B$ will not affect 
%the linear $m-$dependence of equatorially trapped Alfv\'en-Kelvin 
%wave frequencies.}

For small $\tilde k_x\sim 0.18m(M_{33}h_4)^{1/4}(R_6\nu_{3})^{-1/2}$
with $m\neq 0$,
%the frequencies of equatorially trapped Alfv\'en-Rossby waves
%propagating westward are given by
%-----------------------------------------------------------
%---------------------------------------------
%$$
%\tilde\omega\cong -{\tilde k_x\over 2(\tilde k_x^2+n+1/2)}
%$$
%-----------------------------------------------------------
%---------------------------------------------
%
the frequencies of equatorially trapped Alfv\'en-Rossby
waves (lower solid lines of Fig. 1) as seen in an 
inertial reference frame are
$$
f_{\hbox{\eightrm w}}\cong 10^3m\nu_3-
1.3\times 10^2m(M_{33}h_4)^{1/2}R_6^{-2}/(2n+1)\ .\eqno(12)
$$
%$$
%\nu_{\hbox{\eightrm w}}\lsim 10^3m\nu_3+1.5\times 10^2\hbox{ }
%{\nu_3^{1/2}(M_{33}h_4)^{1/4}\over R_6}\qquad (??) \ .
%$$
%this is a visual estimate.
With $m=1$, $n=1$, and a detected $f_{\hbox{\eightrm w}}$, one
can infer $\nu$. For $m=2$ and 3, $f_{\hbox{\eightrm w}}$
would be {\it approximately} 2 and 3 times the fundamental
frequency.

If only one apparent $f_{\hbox{\eightrm w}}$ were detected, 
it would be difficult to distinguish different KARP wave 
possibilities. A detection of two or more harmonics at 2 
and 3 or more times the fundamental frequency 
%with respective amplitudes $\sim 11\%$ and $\sim 4\%$
%of the $401-$Hz amplitude (Wijnands \& van der Klis 1998) 
may offer valuable hints. For nonaxisymmetric equatorially trapped 
Alfv\'en-Poincar\'e waves and mixed Alfv\'en-Rossby-Poincar\'e waves 
traveling around the globe, relation (10) must be modified to include 
the corotation effect $m\nu$ and other $m-$dependence (Fig. 1); the
overall $m-$dependence of these frequencies is not linear. Similarly, the
{\it more accurate} version of relation (12) for Alfv\'en-Rossby waves does
not contain a {\it strictly} linear $m-$dependence either (Fig. 1).
%{\it Given the currently available information,
%(Wijnands \& van der Klis 1998; Chakrabarty \& Morgan 1998),
%equatorially trapped Alfv\'en-Kelvin waves as given by
%equation (22), capable of matching three detected frequencies,
%(Wijnands \& van der Klis 1998)
%would be the most plausible cause of the smooth, nearly sinusoidal
%variation in x-ray brightness of XTE J1808-369 and the inferred
%neutron star spin rate $\nu$ would be $\sim 183$Hz.} As these
%estimates depend on the adopted values of $M_{33}$, $R_6$, $h_4$,
%$C_A$ and $m$, the stellar spin rate $\nu\sim 183$ Hz of XTE J1808-369
%may be modified for a more specific neutron star model as well as
%further observational inferences. 
By detecting more harmonics, one can constrain the type of KARP 
waves involved according to the dispersion relations (Fig. 1). 
As $f_{\hbox{\eightrm w}}$ given by eqns (10)$-$(12) vary with 
$\nu$, the expected stellar $\nu$ spin-up in a LMXB (or unexpected 
variations, such as accreting X-ray pulsar Cen X-3) would lead to 
a $f_{\hbox{\eightrm w}}$ spin-up (or corresponding variations). 

There have been recent theoretical development on stellar $r-$mode 
(Papaloizou \& Pringle 1978; Saio 1982) instabilities caused by 
gravitational radiations (Andersson et al. 1999)
%, Kokkotas, \& Stergioulas 1999) 
to limit spin rates of hot young neutron stars at birth. The prospect 
that fast spinning neutron stars in LMXBs might emit gravitational 
waves (Bildsten 1999) observable by ground-based detectors currently 
under construction (e.g., the enhanced 
%Laser Interferometer Gravitational Wave Observatory 
LIGO) offers another potential test for the KARP wave scenario advanced 
here by independent determinations of neutron star spin rates.

%Mode selection: 10 millions of $p-$modes (acoustic) have been
%observed on the Sun; so far, no sure signals of $g-$modes
%(gravity). $g-$modes have been detected in DA and DB white
%dwarfs with no signals of high-frequency $p-$modes.
%
%These equatorially trapped KARP waves can be also involved
%in kilohertz quasi-periodic oscillations (QPOs) in x-ray
%brightness of LMXBs.

In the case that the stellar surface of a radio 
pulsar is covered with an ocean, various MHD 
tidal waves may modulate or interfere with radio pulse emissions 
from the magnetic polar regions by (quasi-)periodically varying 
polar magnetic field strengths or changing magnetospheric conditions
(Lou 2001a). This may lead to observable effects because magnetic 
and spin axes misalign for radio pulsars to shine. With 
$B\gsim 10^{12}$ G for a pulsar, $C_A$ and $(gh)^{1/2}$ can 
be comparable and be both $\gsim\Omega R$. For equatorially
trapped Alfv\'en-Poincar\'e waves, Alfv\'en-Kelvin waves, and the
high-frequency branch of mixed Alfv\'en-Rossby-Poincar\'e waves,
modulation timescales would be {\it shorter than or comparable to}
radio pulse periods. For equatorially trapped Alfv\'en-Rossby waves 
and the low-frequency branch of mix Alfv\'en-Rossby-Poincar\'e waves,
modulation timescales would be {\it longer than} radio pulse periods. 
These are intriguing possibilities to be explored further.

\acknowledgments

This research was supported in part by grants from US NSF 
(AST-9731623) 
%and the NASA Space Physics Theory and SR\&T Programs through 
to the University of Chicago, 
by the ASCI Center for Astrophysical Thermonuclear Flashes at the
University of Chicago under Department of Energy contract B341495,
by the Visiting Scientist Programs 
at the Institute of Astronomy and Astrophysics, Academia Sinica 
(NSC-88-2816-M-001-0010-6) and at the 
National Center of Theoretical Sciences (Physics Division), 
National Tsinghua University, and by the Collaborative Research 
Fund from the NSF of China for Young Overseas Chinese Scholars 
(NSFC 10028306) at the National Astronomical Observatory, 
Chinese Academy of Sciences.

\clearpage

%% No more than seven \figcaption commands are allowed per page,
%% so if you have more than seven captions, insert a \clearpage

%% There must be a \figcaption command for each legend. Key the text of the
%% legend and the optional \label in curly braces. If you wish, you may
%% include the name of the corresponding figure file in square brackets.
%% The label is for identification purposes only. It will not insert the
%% figures themselves into the document.
%% If you want to include your art in the paper, use \plotone.
%% Refer to the on-line documentation for details.

%% Tables should be submitted one per page, so put a \clearpage before

%% Two options are available to the author for producing tables:  the
%% deluxetable environment provided by the AASTeX package or the LaTeX
%% table environment.  Use of deluxetable is preferred.
%%

%% Three table samples follow, two marked up in the deluxetable environment,
%% one marked up as a LaTeX table.

%% In this first example, note that the \footnotesize command has been
%% used to shrink the table so it will fit on one page. Note also that
%% the \label command needs to be placed inside the \tablecaption.

\clearpage

%% Tables may also be prepared as separate files. See the accompanying
%% sample file table.tex for an example of an external table file.
%% To include an external file in your main document, use the \input
%% command. Uncomment the line below to include table.tex in this
%% sample file.

% \input{table}
%
\begin{figure*}
\vskip 5.0cm
\includegraphics*[angle=0,totalheight=15.5cm]
{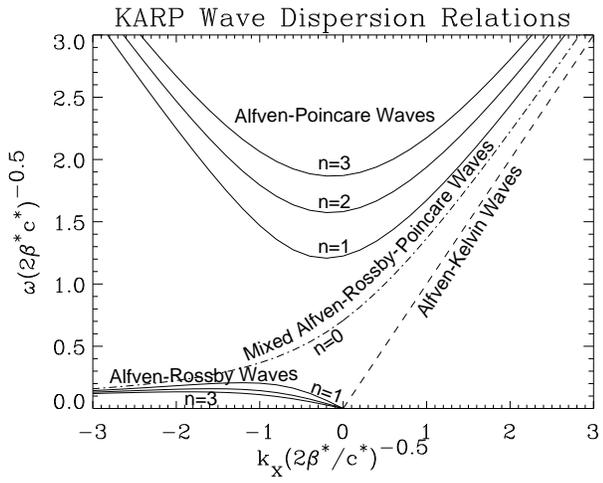}\vskip 0.0cm
\hfill\parbox[b]{17.5cm}{\caption[]{Dispersion relations 
of KARP or MHD tidal waves. \label{f:karp}  } }
\end{figure*}  

%% The following command ends your manuscript. LaTeX will ignore any text

\end{document}